\documentclass[prb,aps,amsmath,amssymb,twocolumn,groupedaddress,floats,showpacs,final,reprint]{revtex4-1}
\usepackage{graphicx}
\usepackage{amssymb,amsmath}
\usepackage{color}
\usepackage{bm}
\begin{document}

%%%%%%%%%%%%%%%%%%%%%%%%%%%%%%%%%%%%%% AUTHORS %%%%%%%%%%%%%%%%%%%%%%%%%
\author{Lode Pollet}
\affiliation{Department of Physics, Arnold Sommerfeld Center for Theoretical Physics and Center for NanoScience, University of Munich, Theresienstrasse 37, 80333 Munich, Germany}
\author{Nikolay V. Prokof'ev}
\affiliation{Department of Physics, University of Massachusetts,
Amherst, MA 01003, USA}
\affiliation{Russian Research Center ``Kurchatov Institute'',
123182 Moscow, Russia}

\author{Boris V. Svistunov}
\affiliation{Department of Physics, University of Massachusetts,
Amherst, MA 01003, USA}
\affiliation{Russian Research Center ``Kurchatov Institute'',
123182 Moscow, Russia}

%%%%%%%%%%%%%%%%%%%%%%%%%%%%%%%%%%%%%%%%%%%%%%%%%%%%%%%%%%%%%%%%%%%%%%%%%%%%%%

%\title{Mesoscopic Classical-Field Criticality of  One-Dimensional Disordered Bosons}
%\title{Classical-Field Flow to Mesoscopic Scales of  One-Dimensional Disordered Bosons}
\title{Classical-Field Renormalization Flow of One-Dimensional Disordered Bosons}

%%%%%%%%%%%%%%%%%%%%%%%%%%%%%%%%%%%%%%%%%%%%%%%%%%%%%%%%%%%%%%%%%%%%%%%%%%%%%%

\date{\today}
\begin{abstract}
We show that in the regime when strong disorder is more relevant than field quantization the superfluid--to--Bose-glass criticality
of one-dimensional bosons is preceded by the prolonged logarithmically slow classical-field
renormalization flow of the superfluid stiffness at mesoscopic scales.
With the system compressibility remaining constant, the quantum
nature of the system manifests itself only in the renormalization of
dilute weak links. On the insulating side, the flow ultimately reaches
a value of the Luttinger parameter at which the instanton--anti-instanton pairs start to proliferate,
in accordance with the universal quantum scenario. This happens first at astronomical system
sizes because of the suppressed instanton fugacity. We illustrate our
result by first-principles simulations.
\end{abstract}


% 03.75.Hh     Static properties of condensates; thermodynamical, statistical, and structural properties
% 67.85.-d     ltracold gases, trapped gases
% 64.70.Tg     Quantum phase transitions
% 05.30.Jp      Boson systems

\maketitle

%%%%%%%%%%%%%%%%%%%%%%%%%%%%%%%%%%%%%%%%%%%%%%%%%%%%%%%%%%%%%%%%%%%%%
\section{Introduction}
\label{sec:1}
%%%%%%%%%%%%%%%%%%%%%%%%%%%%%%%%%%%%%%%%%%%%%%%%%%%%%%%%%%%%%%%%%%%%%

A quarter-century ago, a quantum phase transition from a
superfluid (SF) to Bose glass (BG), a many-body bosonic
counterpart of Anderson localization, was predicted theoretically,
first in one-dimensional (1D) systems by Giamarchi and Schulz (GS)
\cite{GS}, and later in any dimension by Fisher {\it et al.} \cite{Fisher}.
Over the years, substantial progress has been achieved in understanding
this fascinating physics such as describing the interplay between disorder and
commensurability \cite{Weichman}, including the transition to
the so-called Mott glass (MG) phase (rather than BG) in disordered commensurate
systems featuring particle-hole symmetry \cite{Giamarchi_2001},  and the theorem of inclusions, which rules out a direct superfluid--to--Mott-insulator transition and
establishes the Griffiths universality class for transitions out of a
fully gapped state \cite{3d_1,3d_2}. Accurate groundstate phase diagrams of the disordered
bosonic Hubbard model in dimensions $d=2,3$ and partly $d=1$ have been produced by first-principles Monte
Carlo simulations \cite{3d_2,Soyler}, which were also used for large-scale simulations of the universal
critical behavior \cite{large_scale_1,large_scale_2}.
Nowadays, rapid experimental developments in the field of ultra-cold atoms spark renewed excitement~\cite{exp_weak_1,exp_weak_2,exp_weak_3,exp_weak_4,exp_demarco}.

Although many theoretical challenges of disordered bosons have been completed successfully,
there remains one compelling problem in 1D, raised already in Ref.~\onlinecite{GS}: Does a strong-disorder critical point exist which is
qualitatively different from the one found in Ref.~\onlinecite{GS}?  That one was established by means of a perturbative renormalization group (RG) analysis. It was later proved to be a generic scenario for finite disorder strength in Ref.~\onlinecite{instanton} while a recent two-loop RG study\cite{Ristivojevic} is consistent with the latter conclusion. A strong-disorder critical point would also imply the existence of more than one Bose glass phase.

What renders 1D systems special, is the role played by exponentially rare
exponentially weak links (RWLs). Even for a classical-field counterpart of the bosonic system,
weak links can lead to a critical point at which the groundstate loses its macroscopically
uniform response to a phase-twist applied at the boundaries \cite{alexander}.
On the insulating side of this classical transition, a macroscopic system becomes
equivalent to a single Josephson junction with a macroscopically small Josephson coupling.
Inspired by this observation, Altman {\it et al.} proposed the
strong-disorder scenario for quantum systems based on a real-space
RG treatment \cite{Altman2004}.
%In the original version of this scenario~\cite{Altman2004}, quantum
%system was argued to behave similarly to its classical counterpart,
%i.e. at the critical point the superfluid stiffness was gradually
%vanishing with the system size.
As argued in Ref.~\onlinecite{Balabanyan}, the gradual vanishing of the superfluid stiffness down
to zero (corresponding to classical criticality) would be inconsistent for a quantum system since the proliferation of the instanton---anti-instanton pairs is guaranteed in the thermodynamic limit when the Luttinger parameter reaches the critical
%with the exact bound ensuring proliferation of
%instanton--anti-instanton pairs when the Luttinger parameter reaches the critical
value $K_c$ (this phenomenon defines the generic Giamarchi-Schultz critical point \cite{instanton}).
For the SF-BG transition, $K_c=3/2$, while for the SF-MG transition, $K_c=2$, which is the same as for the SF-Mott insulator transition.
Thus, the critical value of the Luttinger parameter in the putative strong-disorder scenario
has to exceed $K_c$~\cite{Altman2004}, which forces the strong-disorder criticality of Kosterlitz-Thouless type
to develop {\it prior} to the generic mechanism of Refs.~\onlinecite{GS, instanton}.

\begin{figure}[tpb]
\includegraphics[width=1.0\linewidth]{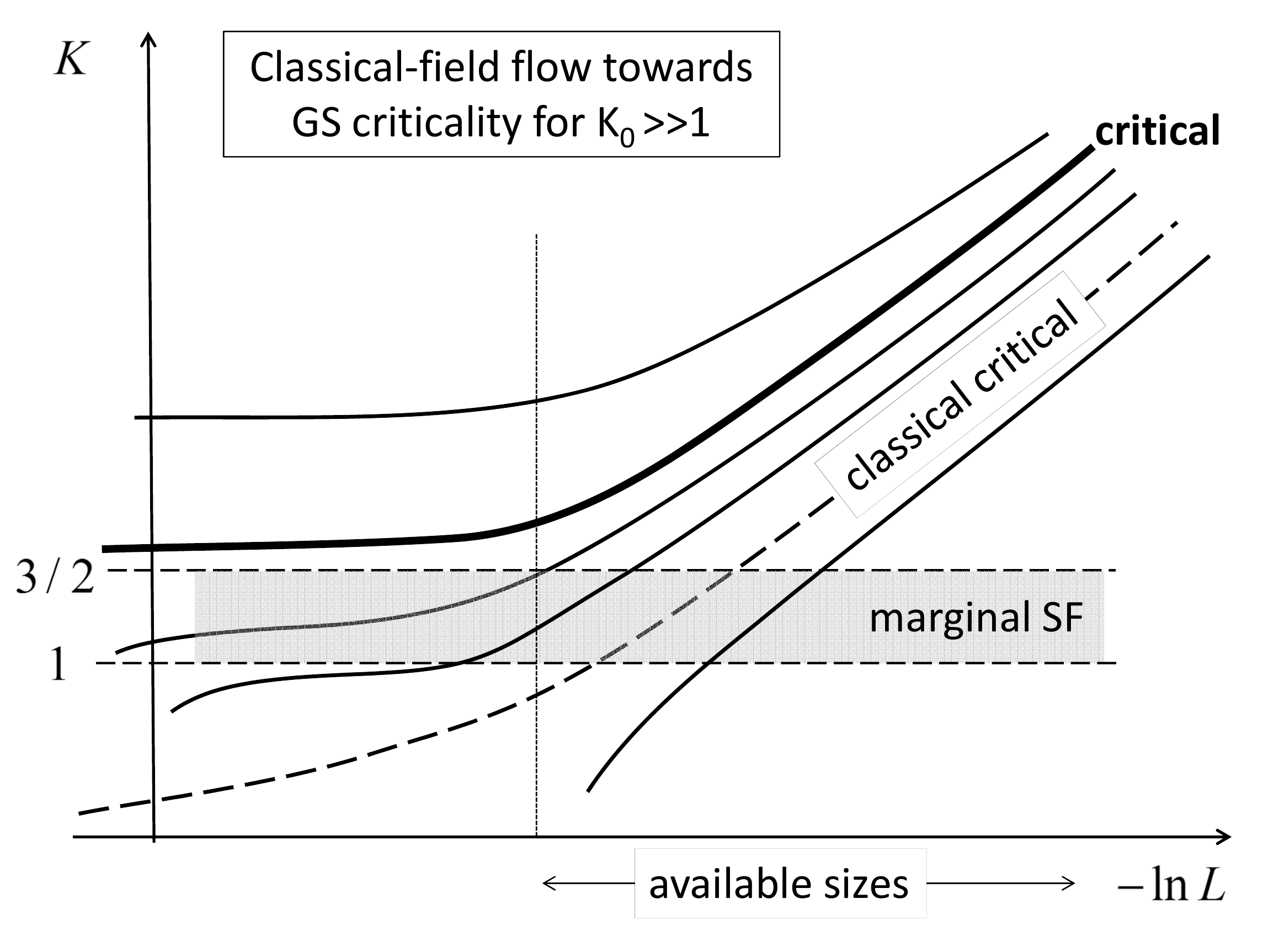}
\caption{\label{fig:GS}
Renormalization group flow of the Luttinger parameter which starts as classical
flow and crosses over to the standard Kosterlitz-Thouless flow at much larger scales
if the former saturates to a value slightly above or below $K_c=3/2$. These large scales
may not be available for numerical analysis.
}
\end{figure}

\begin{figure}[tpb]
\includegraphics[width=1.0\linewidth]{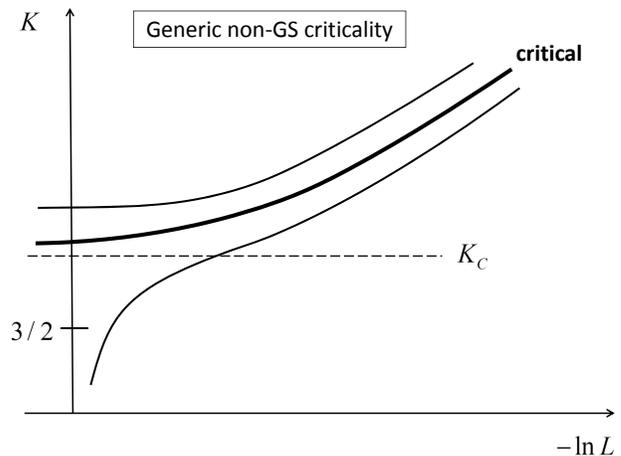}
\caption{\label{fig:KT}
Single-stage renormalization group flow in the hypothetical strong-coupling
scenario. Having passed the correlation length $r_0$ corresponding to the condition $K(L=r_0) = K_c> 3/2$, the flow
rapidly brings the system to localization. }
\end{figure}

In this paper, we show that the RG flow of $K(L)$ can retain its classical-field behavior
down to the $K(\infty )=K_c$ value, where the universal phenomenon of
instanton proliferation inevitably takes place on sufficiently large
scales. This scenario occurs when $K(\xi_0)=K_0\gg 1$ at the microscopic cutoff scale $\xi_0$, {\it i.e.}, in the most natural regime for the putative strong-disorder scenario. The quantum
renormalization of the RWLs is irrelevant for the flow behavior.
In this limit, not only the putative strong-disorder critical point is ruled out,
but also the generic instanton-pair induced flow is practically inoperational because
of the extremely low fugacity of  pairs of mesoscopic size.
As a result, a typical system of linear size $L$ that should formally be called
an insulator, because $K(L)<K_c$,  retains a superfluid response. This behavior (illustrated in  Fig.~\ref{fig:GS})---as we will later see, this is the hallmark of our scenario,
distinguishing it from both the Giamarchi-Schulz case and the quantum strong-disorder regime advocated in Ref.~\onlinecite{Altman2004}---persists over a broad range of scales till $K(L)\approx 1$ (see further on in the text). For the same reason of extremely low instanton fugacity the critical RG flow features
a pronounced two-stage character (see Fig.~\ref{fig:GS}, and note that the accessible sizes may be too small to observe both stages of the flow), as opposed to the single-stage flow of the hypothetical quantum
strong-disorder regime (see Fig.~\ref{fig:KT}).

We also discuss that a strong-disorder scenario with the critical value of $K_c$ significantly above
the GS value may not look quantitatively similar to Fig.~\ref{fig:GS} because if the classical flow
levels off at $K$ significantly above the $3/2$ value, the proliferation of instantons with small fugacity
(or backscattering processes) becomes impossible. This implies that any strong-coupling scenario
with $K>3/2$ should have a single-stage RG flow illustrated in Fig.~\ref{fig:KT}. We also
prove that any quantum critical flow possesses self-averaging properties and thus is subject to
asymptotic hydrodynamic description. As far as we see, this cannot be reconciled with any strong-coupling scenario.

While the condition $K_0\gg 1$ is necessary to guarantee the applicability of
our theory, in practice, the mesoscopic classical-field strong-disorder
flow can take place already at $K_0 \sim 3$, which we clearly see
in first-principles simulations of a J-current type model up to
system sizes $L\sim 1000$.

The rest of the paper is organized as follows. In Sec.~\ref{sec:2} we develop our theory.
In Sec.~\ref{sec:3} we address the subtle issues of self-averaging under the conditions of strong disorder,
arguing that the observables of choice are the median values and appropriately defined widths rather
than the standard cumulants---the expectation value and variance, or the dispersion which is the square root of the variance. In Sec.~\ref{sec:4} we present the model and the results of an illustrative
simulation. Our conclusions are formulated in Sec.~\ref{sec:5}.

%%%%%%%%%%%%%%%%%%%%
\section{Theoretical Analysis}
\label{sec:2}
%%%%%%%%%%%%%%%%%%%%%

The main assumption---to be justified later---of the instanton approach
\cite{instanton} is that up to the critical point the system remains a
well-defined Luttinger liquid characterized by a finite parameter
$K$. This allows one to describe the coarse-grained behavior of
the system with Popov's hydrodynamic action $S[\Phi]$, where
$\Phi\equiv \Phi(x,y)$ is the $(1+1)$-dimensional field of
superfluid phase ($x$ is the spatial coordinate and $y=c\tau$ is
the imaginary time re-scaled with the sound velocity $c$). The
instantons are the virtual quantum events which result in the phase slips
of $\Phi$. In $(1+1)$ dimensions they are vortices in the field $\Phi(x,y)$.
An important distinction between the instanton  and  its 2D vortex counterpart
is that the contribution of an instanton to the partition-function integrand
comes with the $x$-dependent factor ${\rm e}^{ \pm i \gamma(x)}$, in which the sign
is defined by the sense of the vortex rotation and $\gamma(x)=2\pi \int_0^x n(x')\, dx'$,
where $n(x)$ is the disorder-dependent microscopic expectation
value of the number density at point $x$.  In view of the
random phase factor $\gamma(x)$, the net contribution
of instantons to the partition function survives after integration
over $x$ only if instantons come in instanton--anti-instanton
pairs with two $x$-coordinates being microscopically close to each other
to ensure that their phase factors compensate each other.
[This is not the case for the MG where exact particle-hole symmetry and commensurability
render $n(x)=integer$ and the phase factor irrelevant. It is precisely this aspect that
leads to the difference between the $K_c$ values for BG and MG transitions.]
For our purposes, we do not need to consider the general case of RG flow
of $K(L)$ due to the instanton--anti-instanton (or vortex-antivortex, for
briefness) pairs. When the gas of pairs is extremely dilute,
the criterion for their proliferation is derived based on the statistics
of a single pair with a hierarchy of length scales starting from the
microscopic cutoff $\xi_0$ to the system size $L$ being absorbed into the scale
dependent value of $K(r)$. The proliferation takes place when the ``effective energy" of the pair
(in effective temperature units, $T=1$)
\begin{equation}
E_{\rm pair} = 2\int_{\xi_0}^L \frac{K(r)}{r}\, dr + E_c,
\label{eq:vortex_energy}
\end{equation}
with $E_c$ the (unknown) vortex pair core energy, becomes
lower than the entropy given by its three degrees of freedom: two in
the $y$-direction and one in the $x$ direction (there are 4 degrees of freedom for the MG case)
\begin{equation}
S_{\rm BG} = 3 \ln L , \qquad S_{\rm MG} = 4 \ln L,
\label{eq:vortex_entropy}
\end{equation}
so that the effective free energy turns negative. In the genuine
thermodynamic limit of $L\to \infty$, this happens at the
universal value $K(\infty)=K_c$. For $K(\xi_0)\gg 1$
and given the slow decrease of $K$ with distance, an enormously
large system size is required for the entropy term to compete with
the contribution in $E_{\rm pair}$ coming from distances where
$K(r)\gg 1$. This explains the extremely low fugacity of the
vortex-antivortex pairs and accounts for the fact that
$E_{\rm pair}(L)$ can remain larger than $S(L)$ even when
$K(L)$ is already substantially {\it lower} than $K_c$.

To force a system with large $K_0$ ending up in the insulating state
one needs strong disorder which produces enough weak links to renormalize $K$
down to $K_c$. For the flow to have classical-field character, the microscopic
physics of RWL does not necessarily have to be classical.
Quantum effects can dramatically change the actual strength of the link
by dressing it with vortex-antivortex pairs sitting {\it right at the link};
the effective energy of such pairs can be quite small
[note that a space-time isotropic Eq.~(\ref{eq:vortex_energy}) does not apply to RWLs].
However, in accordance with the result by Kane and Fisher \cite{Kane-Fisher}
(see also Ref.~\onlinecite{instanton} for the instanton interpretation)
the macroscopic proliferation of vortex-antivortex pairs on a link, no matter how weak,
takes place only at $K<1$ (in the thermodynamic limit). Hence, as long as $K>1$,
the quantum physics of RWLs can be absorbed into their renormalized Josephson couplings, a circumstance which we rely upon in our analysis,
which consequently closely follows ideas developed in Ref.~\onlinecite{alexander}.

To proceed, we have to specify the distribution of RWLs.
Here we confine ourselves with a (rather typical) strong-disorder setup when
RWL emerges as a sequence of $\ell_*$ statistically independent  microscopic ``insulating" regions,
in which case the probability of finding the RWL somewhere in the system of the (dimensionless) size $L$ is given by
\begin{equation}
\rm{Prob} [ \ell_*,L ] \, \propto\,  (L/ \ell_*)\, {\rm e}^{- c_1 \ell_* } \;,
\label{eq:prob}
\end{equation}
where $e^{-c_1}$ defines the probability of a single ``insulating" region to occur at a given point.
We are interested in the case when this probability is of order unity, from which the dependence
of $\ell_*$ on system size $L$ to logarithmic accuracy follows,
\begin{equation}
\ell_* \approx c_1^{-1} \left( \ln L - \ln \ln L  \dots \right).
\end{equation}
The second term is related to the Kinchin-Kolmogorov law of the iterated logarithm.
%which states how the sum of $N$ outcomes in a random walk fluctuates in magnitude around  $\Delta \sqrt{N}$.
The Josephson coupling $J$ over the link is
\begin{equation}
J(\ell_*) =  F(\ell_*)\,  {\rm e}^{-c_2 \ell_*} \approx  F( \ln L) \,
( \ln L /L )^{c_2/c_1} \;,
% \left( \frac{\ln L}{L} \right)^{c_2/c_1} \;,
% {\rm e}^{ - (c_2/c_1) (\ln L - \ln \ln L)} ,
\label{eq:Josephson}
\end{equation}
where the function $F(\ln L)$  and the constant  $c_2$ absorb all the microscopic physics of the weak link.
The inverse of the superfluid stiffness $n_s(L)$ for the system of size $L\gg \xi_0$ is then (classically)
given by the sum of inverse Josephson couplings of RWLs,
\begin{equation}
n_s^{-1}(L) = n_s^{-1}(\xi_0) +  \left( J_1^{-1} +  J_2^{-1}+ J_3^{-1}\ldots \right)/L\, .
\label{eq:semiclass}
\end{equation}
As it will be seen self-consistently, the nature of the flow at the classical critical point,
implied by Eqs.~(\ref{eq:Josephson}) and~(\ref{eq:semiclass}), is such that doubling the system
size leads only to a minor change in $n_s(L)$, that is the $J$ value of the weakest link
gets {\it progressively larger} than $n_s(L)/L$ as $L\to \infty$.
Hence, a substantial change of $n_s(L)$ happens over a broad range of length scales
involving a large number of RWLs at different length scales.
This  allows one to treat $n_s(L)$ as a slow continuous function of $L$ and
replace Eqs.~(\ref{eq:Josephson})-(\ref{eq:semiclass}) with the flow equation
\begin{equation}
\frac{d\, n_s^{-1}} {d \ln L} \, =  \,  \tilde{F}(\ln L) \, L^\zeta \,  +\,  {\rm O}( \ln \ln L) \, ,
\label{eq:flow_ns}
\end{equation}
where $\zeta = c_2 /c_1  - 1$. Note that the correction terms are iterated logarithms in the system size.
By the same token, the flow of the compressibility with $\ln L$ is leaving it constant to this order of accuracy.

The critical condition corresponds to $\zeta \to 0$. At this point $n_s(L)\equiv n_s(\ln L)$ is decaying
logarithmically slowly justifying self-consistently the assumption of being a smooth function of $L$.
For $|\zeta | \ll 1$, a substantial part of the RG flow remains slow and smooth even on the insulating side $\zeta>0$.

The flow (\ref{eq:flow_ns}) predicts  $K(\zeta=0,\, L=\infty)= 0$.
Quantum effects do not change this flow even for $K<K_c$ until $L$ is large enough for
free instantons to appear. The actual quantum critical line corresponds to the regime when the
classical critical flow takes the system to $K(L=\infty)=K_c$. If $K_0\gg1$, one can reveal how the
classical-field flow ultimately crosses over to the universal quantum one
only at astronomically large $L$ leaving  no chances to observe it in a cold gas experiment nor in first-principles simulations.
What can be observed, however,
is the anomalous finite-size behavior of the superfluid stiffness. On the superfluid side ($\zeta < 0$), finite size corrections
$n_s(L) - n_s(\infty) \propto  1/L^{|\zeta|}$ can, in principle,  be used to extract the critical parameters
of the Hamiltonian.

On the insulating side, a typical feature is the development of substantial sample-to-sample fluctuations of $n_s(L)$
reflecting the growing effect of single RWLs on $n_s(L)$. Somewhat away from the critical condition $\zeta=0$
(still considering $\zeta\ll 1$) RWLs ultimately destroy the self-averaging of $n_s(L)$ [and thus applicability of Eq.~(\ref{eq:flow_ns})]
at large enough $L$.
One can imagine that in a peculiar regime with condition $0<\zeta \ll 1$  the self-averaging of $n_s(L)$ survives till $K(L=L_*) = 1$, while the suppressed fugacity of
instantons prevents the system from having insulating properties.
In such a case, the Kane-Fisher physics on a single RWL will be responsible for rendering the
system insulating at  $L>L_*$.

The theory developed in this section can be used for processing numeric and/or experimental data,   allowing
one to predict the behavior of the system in question at much larger sizes and, in particular, extract critical parameters of disorder.
A practical example will be considered in Sec.~\ref{sec:4}, preceded by Sec.~\ref{sec:3} in which we address delicate aspects of self-averaging
characteristic of the strong-disorder classical-field flow, and formulate appropriate statistical observables to be associated with (marginally) self-averaged quantities
of the present section.

%%%%%%%%%%%%%%%%%%%%%%%%%%%%%%%%%%%
\section{Statistical Subtleties. Theorem of Critical Self-Averaging}
\label{sec:3}
%%%%%%%%%%%%%%%%%%%%%%%%%%%%%%%%%%

Introduction of the notion of superfluid stiffness as an intensive quantity requires self-averaging in the superfluid phase,
at least in a certain minimalistic sense requiring that relative fluctuations of $n_s(L)$ among different disorder realizations be small and get progressively smaller with increasing $L$ (see also further on in this section and in Sec.\ref{sec:4} for a numerical estimate of the minimum system size required for this to hold in the superfluid phase).
The same is true for $n_s^{-1}(L)$, the prime object of the analysis of the previous section.
Nevertheless, at small $\zeta$, the issue of self-averaging becomes  delicate because
of the potential divergence of the {\it dispersion} of $n_s^{-1}(L)$ in the $L \to \infty$ limit \cite{Polkovnikov}.
[This divergence does not necessarily contradict with a vanishing characteristic width of the distribution.]

While not creating fundamental theoretical difficulties, the divergence of dispersion requires some care in processing finite-size data. Obviously, one has to avoid working with the
dispersion when characterizing fluctuations of $n_s^{-1}(L)$.  Correspondingly, in order to characterize the distribution of a random variable $x$ we use three typical numbers (below $P$ stands for probability): (i) the median $\bar{x}$ such that $P[x<\bar{x}]=1/2$, (ii) the lower characteristic value $x_-$ such that  $P[x<x_-]=1/4$, and (iii) the upper characteristic value $x_+$ such that  $P[x>x_+]=1/4$. The width of the distribution is then naturally characterized as $\delta x= x_+ - x_-$.
Correspondingly, by the relative width we understand $R=\delta x/\bar{x}$.

A special convenience of working with $x_{\pm}$ and $\bar{x}$ is due to the fact that for any random number $y=f(x)$ related to $x$ by a monotonic function $f$,  we have $\bar{y}=f(\bar{x})$ and $y_{\pm}=f(x_{\pm})$, or $y_{\pm}=f(x_{\mp})$.
In what follows all physical quantities are characterized by their medians and (relative) widths, {\it not} the first and the
second moments of the distribution functions.

We now prove a theorem that self-averaging in the above mentioned sense takes place at the
critical line separating SF and BG phases, no matter whether it is of the GS type, or corresponds to some
hypothetical strong-coupling scenario based on weak links, which features a critical value of the Luttinger
liquid parameter $K_c >3/2$. Finite $K_c$ implies finite critical value of $1/n_s$ because
with the compressibility remaining finite at the critical point we can speak of $K$ and
$1/n_s$ interchangeably at the level of their medians. This is consistent with the most conservative
definition of the superfluid phase and the critical point at its boundary
as states with non-zero probability density, $P^{(L)}(x)$, of finding a finite value of $x \equiv 1/n_s$ in the thermodynamic limit $L\to \infty$.
This leads to a meaningful (i.e. with non-zero weight over a large-enough finite interval)
limiting function which is allowed---but not {\it a priori} required---to be a $\delta$-function:
\begin{equation}
\tilde{P}(x) = \lim_{L\to \infty} P^{(L)}(x) \, .
\label{P_lim}
\end{equation}

Consider now the process of combining two macroscopically large systems of size $L$
into one system of size $2L$ by adding a link between their edges. Due to the local renormalization
of the weak link physics we accept that the contribution of the link to the macroscopic
stiffness may depend on properties of the left and right pieces, but not on their macroscopic size.
In other words, if on the left the superfluid stiffness is  $x$ and on the right it is $y$,
then the link distribution function is $Q_{x,y}(\xi=1/J)$ and the composition law for the
probability distribution of $z=1/n_s$ in the combined system is [see (\ref{eq:semiclass})]
\begin{eqnarray}
P^{(2L)}(z) \, = \, \int_0^{\infty}dx  \int_0^{\infty}dy  \int_0^{\infty}d\xi \qquad   \qquad   \qquad   \qquad  \nonumber \\
P^{(L)}(x)\, P^{(L)}(y)\, Q_{x,y}(\xi) \, \delta \! \left [z\! -\! {x\! +\! y\! +\! \xi/L \over 2}\right]  .  \qquad
\label{composition1}
\end{eqnarray}
The distribution of weak links is certainly normalizable. Moreover, even at classical criticality
it decays at large $\xi$ as a Lorentzian. Since, at any fixed $z$, the ranges of integrations
in (\ref{composition1}) are finite, we can safely take the limit of $L\to \infty$ and
replace $P$ with $\tilde{P}$
\begin{eqnarray}
\tilde{P}(z) \, = \, \int_0^{\infty}dx  \int_0^{\infty}dy  \int_0^{\infty}d\xi \qquad   \qquad   \qquad   \qquad  \nonumber \\
\,  \tilde{P}(x)\, \tilde{P}(y)\, Q_{x,y}(\xi ) \, \delta \! \left [z\! -\! {x\! +\! y\! +\! \xi /L \over 2}\right]  . \quad
\label{composition2}
\end{eqnarray}
Integration over $z$ in Eq.~(\ref{composition2}) then yields
\begin{equation}
f\, =\, f^2 q  \, ,
\label{f_q}
\end{equation}
where
\begin{equation}
f= \int_0^{\infty} \tilde{P}(x) \, dx \, \leq 1\, ,~ \quad q= \int_0^{\infty} \tilde{Q}(\xi) \, d\xi \, = 1\, .
\label{f_q_int}
\end{equation}
Since $f>0$ (without a finite fraction of systems with non-zero $n_s$, we fail to meet the definition
of the critical state with finite $n_s$), from Eq.~(\ref{f_q}) we immediately conclude that
$f=1$, meaning that at the critical point the limiting distribution $\tilde{P}(x)$ is normalized to unity.

In Fourier space, Eq.~(\ref{composition2}) reads
\begin{equation}
\tilde{P}_k=\! \int \! dx\, dy\,  \tilde{P}(x)\tilde{P}(y){\rm e}^{ik(x+y)/2}
\!    \int \! d\xi \, {\rm e}^{ik\xi/2L} \, Q_{x,y}(\xi)  \, .
\label{Fourier}
\end{equation}
Since the last integral corresponds to macroscopically small Fourier harmonics of $Q_{x,y}$  that
decay at large $\xi$ at least as fast as a Lorentzian, we can replace it with unity [the leading correction
is $\mathcal{O}(1/L)$]. This leaves us with the equation
\begin{equation}
\tilde{P}_k = \tilde{P}_{k/2}^2 \;,
\label{PP2k2}
\end{equation}
for which we have two options: (i) Either $ |\tilde{P}_{k} |\equiv 1$, or (ii) $|\tilde{P}_k|$ decays exponentially
at $|k|\to \infty.$ Option (ii) is incompatible with the requirement that the function
$\tilde{P}(x)$ be  identically zero at $x< 0$. Hence, we have $\tilde{P}_k={\rm e}^{ik/\bar{n}_s}$, i.e.
$\tilde{P}(x)=\delta(x-1/\bar{n}_s)$, which  proves the theorem.

Two remarks are in order to prevent possible confusions. Having observed that typical values of $J$ in the composition law (\ref{composition1}) are of order unity, while
the strength of characteristic weak links responsible for, say, the classical flow (\ref{eq:flow_ns}) is macroscopically small, one may wonder how the two facts should be reconciled
with each other, and, in particular, how those macroscopically small links are properly taken into account 
in the composition law (\ref{composition1}). The first confusion is removed by noting that the weak links in Eq.~(\ref{eq:flow_ns}) are being selected throughout the whole system, while the link in the composition law (\ref{composition1}) is found at a given
position in space, right at the center of the system. The macroscopic difference in two procedures thus 
accounts for the macroscopic difference in characteristic values of the corresponding links. 
The role of macroscopically weak links is implicit in Eq.~(\ref{composition1}) --- those links effect the form of the distribution $P^{(L)}(x)$.

Our second remark concerns confusing self-averaging with the existence of a finite dispersion. One should not think that $\tilde{P}(x)=\delta(x-1/\bar{n}_s)$ somehow implies
a finite dispersion at any finite $L$.  The dispersion depends on the limit taken to obtain the generalized function.
Self-similar distributions for random variables based on sums of identical random variables
are known in mathematics as {\it stable} distributions.\cite{Feller1, Feller2}
A distribution for a random variable $X$ is said to be stable if for two independent copies $X_1$ and $X_2$  and for any constants $a >0,b >0$ it holds that $aX_1 + bX_2 = \gamma X + d$ for some constants $\gamma > 0$ and $d$.
Normal distributions are stable distributions (by the central limit theorem, when the variance is finite),
but there exists a {\it generalized central limit theorem}
which states that the sum of a number of random variables with power-law tail distributions
decreasing as $|x|^{-\alpha-1}$ where $0 < \alpha < 2$ (and therefore having infinite variance)
will tend to a stable distribution of the from $P(x) = \frac{1}{\pi} \int {\rm e}^{itx} {\rm e}^{-(\gamma t)^{\alpha} }dt$, with scale parameter $\gamma > 0$ which is a measure for the width.
Note that this goes beyond the law of the large numbers. The distribution can be centered around $\mu$ by the substitution $x \to x - \mu$.
In particular, for $\alpha = 1$ we have the well-known Lorentz distribution which has infinite moments at any finite $\gamma$.
Physically, we know that the divergence of the dispersion with $L$ is certainly happening for a marginal superfluid,
but the system will ultimately flow to an insulator.
However, one can also envisage situations in a superfluid where the bi-modality due to the divergence of
the dispersion is resolved by keeping the location of the center $\mu$ finite while the statistical importance
of the tail is irrelevant because of macroscopical pre-factors suppressing the tail (cf. the stable distributions above).
Such examples should not be thought of as a different superfluid phase (from the usual one with vanishing dispersion at any finite $L$) because
self-averaging does take place in the thermodynamic limit with only rare, thermodynamically unimportant fluctuations.

\begin{figure}[t]
\includegraphics[width=1.02\linewidth]{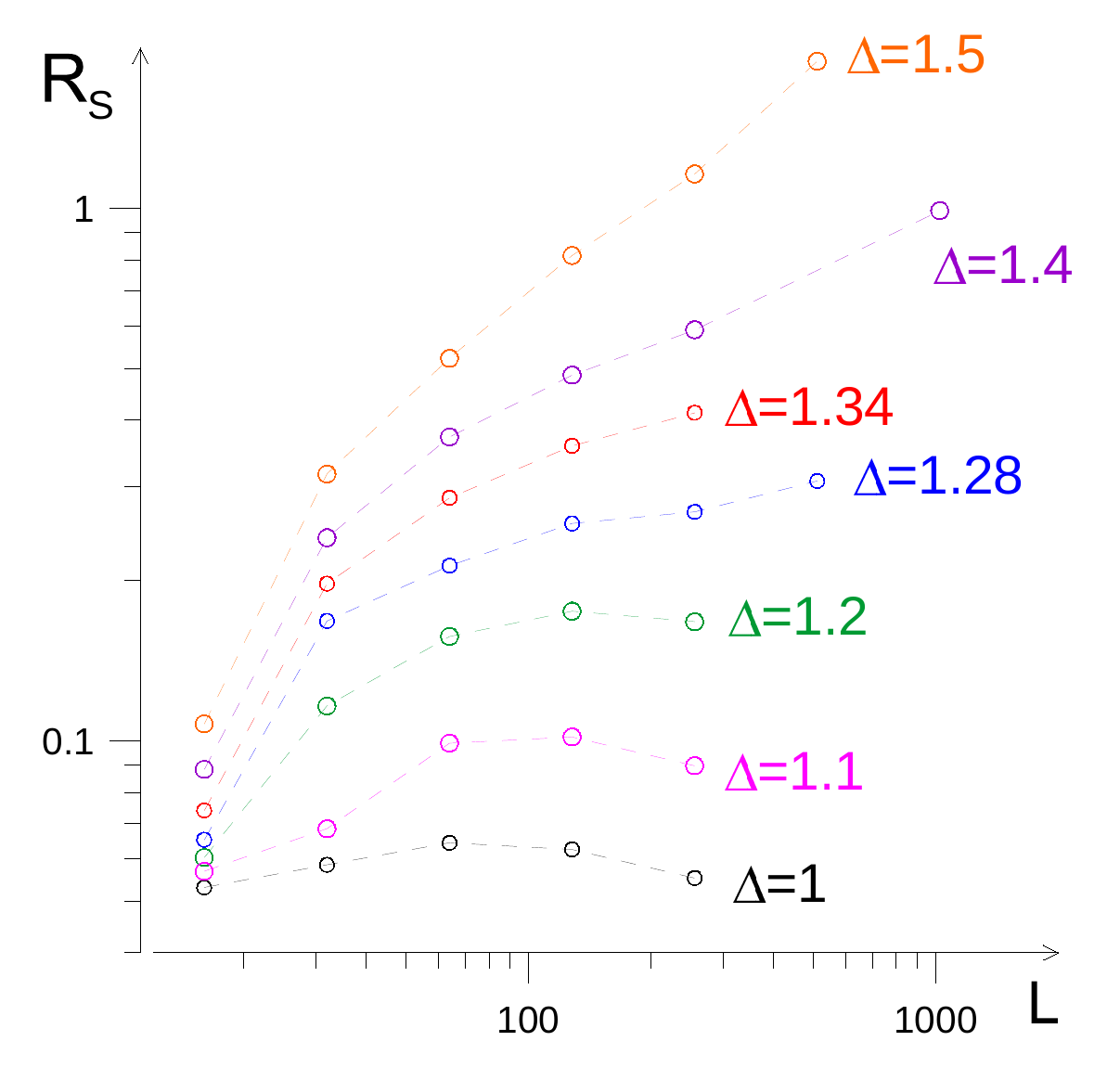}
\caption{\label{fig:width}(Color online). The relative width  $R_s = \overline{n}_s  \, \delta n_s^{-1}$ of the distribution of $n_s^{-1}$ (calculated in accordance with conventions of Sec.~\ref{sec:3}). The increase in relative width for $\Delta \geq1.28$
leads to insulating behavior, according to the theorem of self-averaging on the critical line.
}
\end{figure}

The theorem we proved can, in particular,  be used to identify
which systems flow to insulators (cf. Fig~\ref{fig:width}).
Moreover, self-averaging on the critical line allows one to use the hydrodynamic description at large distances.
By continuity, even if one increases disorder strength beyond the critical value by some arbitrary small
but finite amount, $\Delta = \Delta_c + \epsilon$, the hydrodynamic description will apply up to
enormously large distance scales. Next we notice (by the instanton analysis described above),
that if this is taking place with $K_c>K > 3/2$, then disorder fluctuations become less and less relevant
with increasing the distance scale, {\it i.e.}, the state will retain its superfluid properties and will
not become an insulator, in formal contradiction with the $K_c > 3/2$ scenario. We do not see any
physical mechanism which invalidates this conclusion.

%%%%%%%%%%%%%%%%%%%%%%%%%%%%%%%%%%%
\section{Illustrative Simulation }
\label{sec:4}
%%%%%%%%%%%%%%%%%%%%%%%%%%%%%%%%%%

%%%%%%%%%%%%%%%%%%%%%%%%%%%%%%%%%%%%%%%%%%%%%%%%%
\begin{figure}[tpb]
\includegraphics[width=1.0\linewidth]{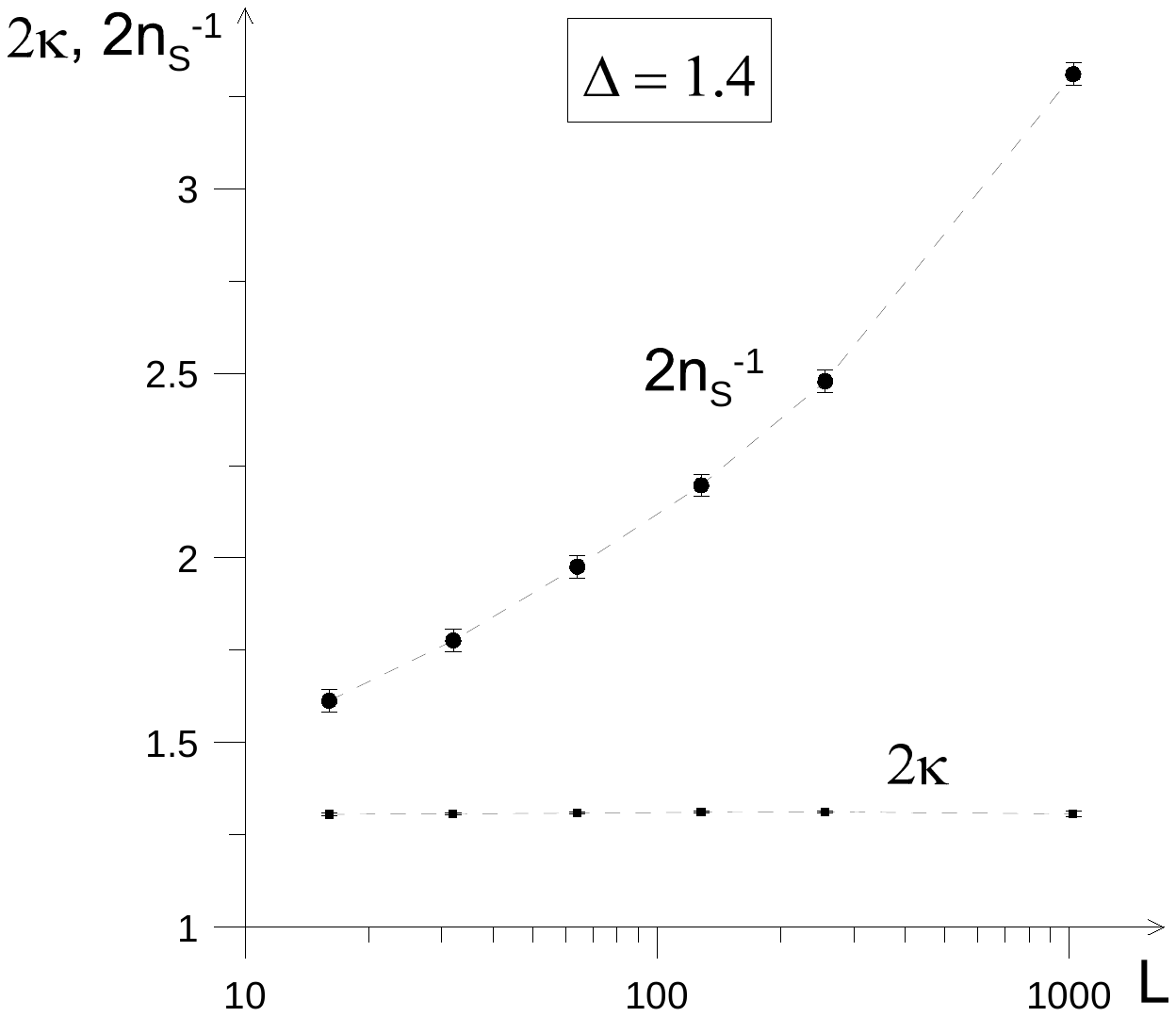}
\caption{\label{fig:sf_kappa}(Color online). Median inverse stiffness, $n_s^{-1}$, and compressibility, $\kappa$,
as functions of system size $L$ for disorder strength $\Delta = 1.4$. (Compressibility error bars are shown but  too small to be resolved in this plot.)
}
\end{figure}
%%%%%%%%%%%%%%%%%%%%%%%%%%%%%%%%%%%%%%%%%%%%%%%%%

%%%%%%%%%%%%%%%%%%%%%%%%%%%%%%%%%%%%%%%%%%%%%%%%
%%%%%%%%%%%%%%%%%%%%%%%%%%%%%%%%%%%%%%%%%%%%%%%%%

%%%%%%%%%%%%%%%%%%%%%%%%%%%%%%%%%%%%%%%%%%%%%%%%%
\begin{figure}[t]
\includegraphics[width=1.02\linewidth]{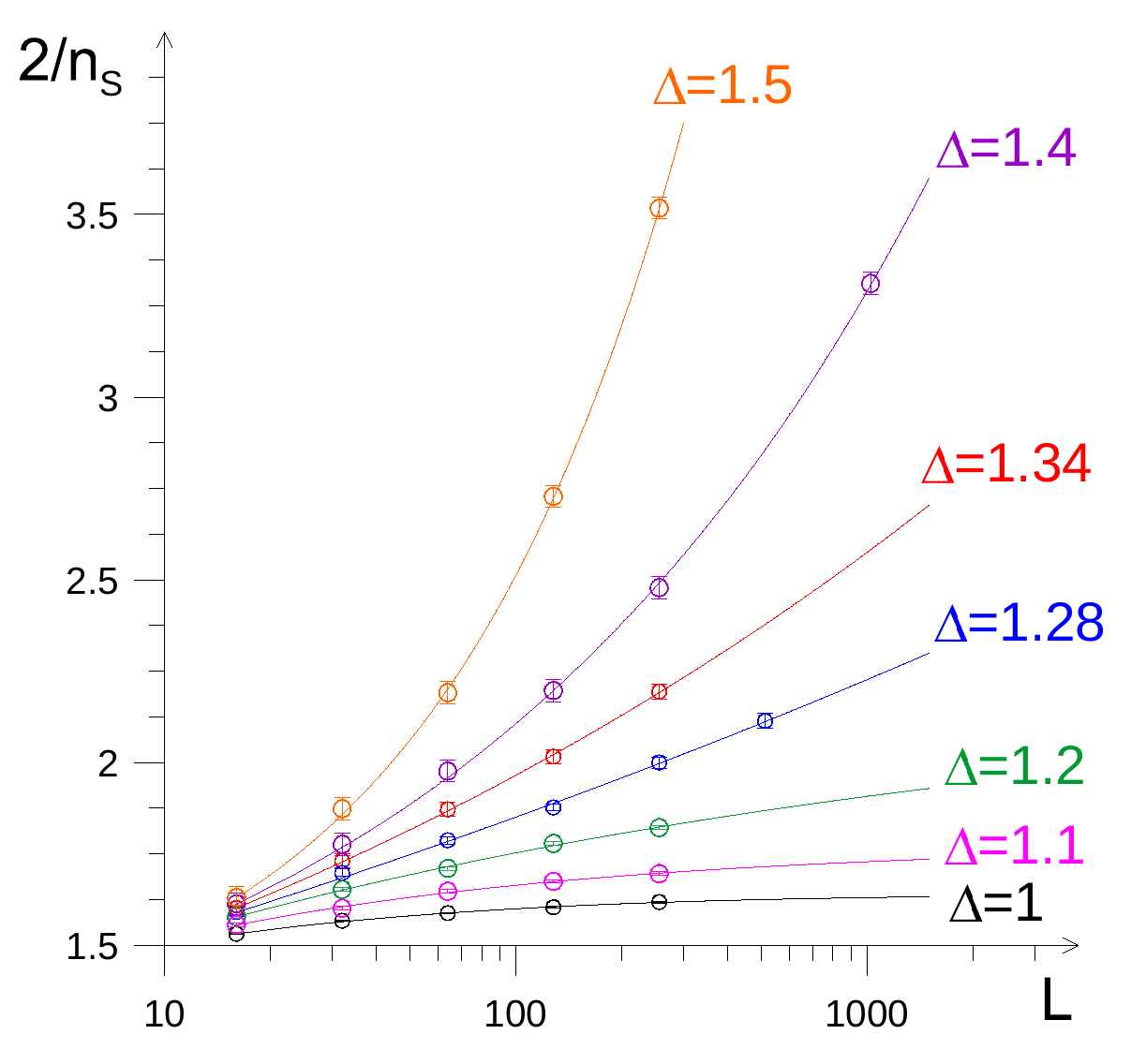}
\caption{\label{fig:glory}(Color online). The flows of the median values of $n_s^{-1}(L)$ fitted with Eq.~(\ref{eq:flow_ns}) in which the (unknown) function
$\tilde{F}(\ln L)$ is chosen to be a ($\Delta$-dependent) constant, $\tilde{F}\equiv B(\Delta)$, treated as a fitting parameter (see also Fig.~\ref{fig:B}).
}
\end{figure}
%%%%%%%%%%%%%%%%%%%%%%%%%%%%%%%%%%%%%%%%%%%%%%%%%

%%%%%%%%%%%%%%%%%%%%%%%%%%%%%%%%%%%%%%%%%%%%%%%%%
\begin{figure}[t]
\includegraphics[width=1.02\linewidth]{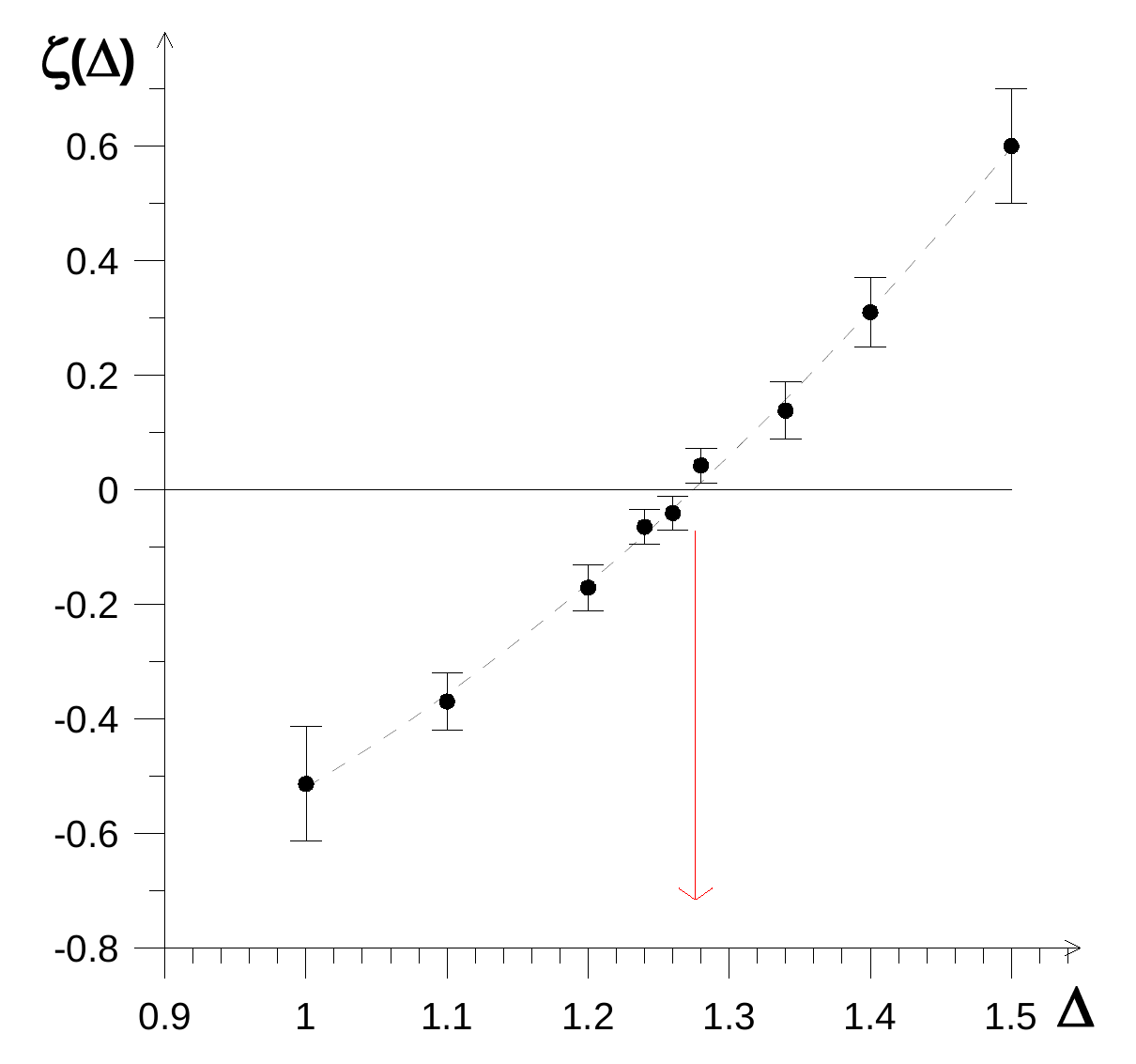}
\caption{\label{fig:zeta} The exponent $\zeta(\Delta)$. The arrow indicates the classical critical value of disorder, $\Delta_c^{\rm (cl)} =1.275(10)$,
such that $\zeta\left( \Delta_c^{\rm (cl)} \right) =0$. The function $\zeta(\Delta)$ is structureless in the vicinity of $\Delta=\Delta_c^{\rm (cl)} $.}
\end{figure}
%%%%%%%%%%%%%%%%%%%%%%%%%%%%%%%%%%%%%%%%%%%%%%%%%

%%%%%%%%%%%%%%%%%%%%%%%%%%%%%%%%%%%%%%%%%%%%%%%%%
\begin{figure}[t]
\includegraphics[width=1.02\linewidth]{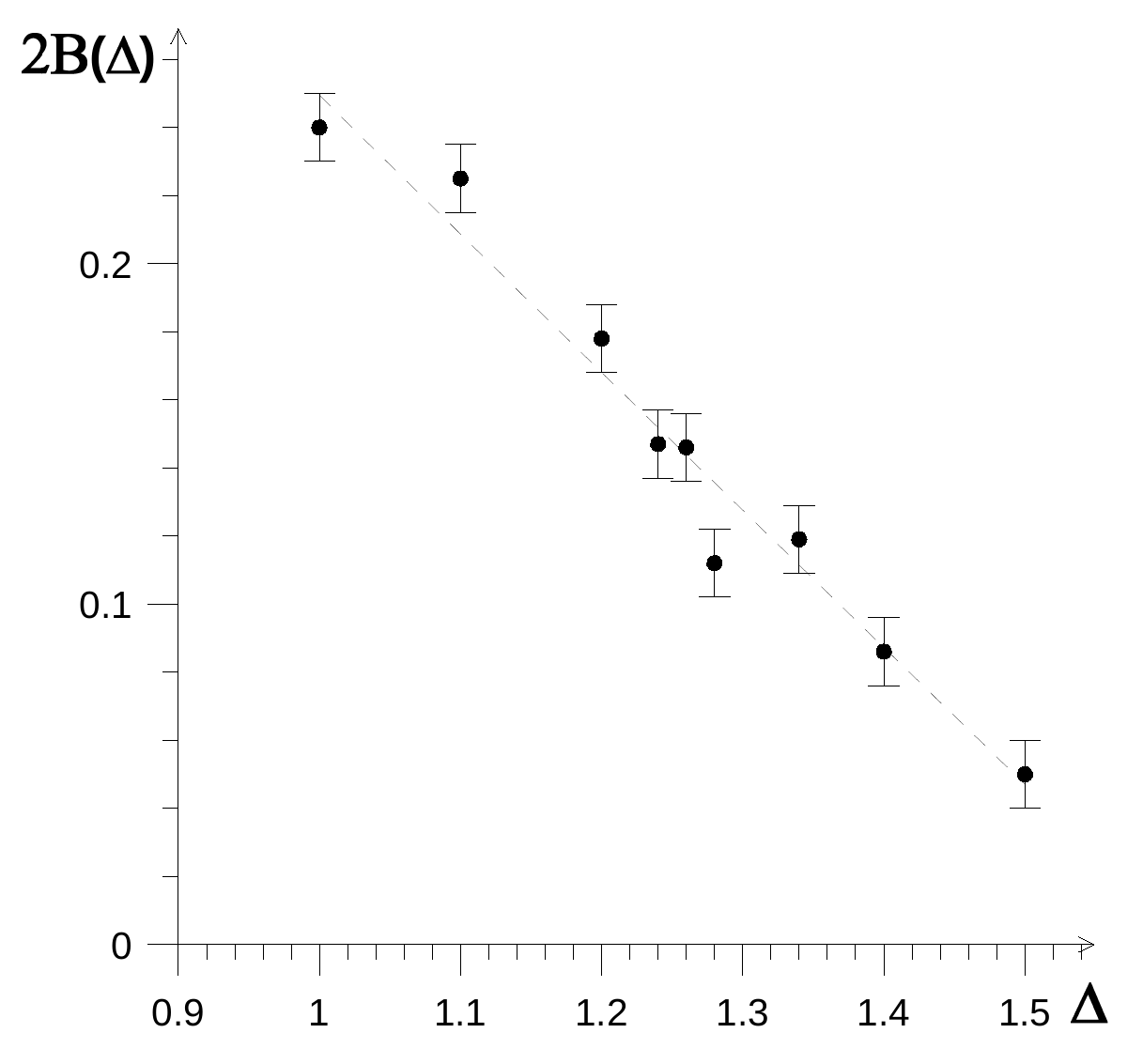}
\caption{\label{fig:B} The  parameter $\tilde{F}\equiv B(\Delta)$ extracted from the fits. Within the error bars, the dependence of $B$ on $\Delta$
is linear, as shown by the dashed line.}
\end{figure}
%%%%%%%%%%%%%%%%%%%%%%%%%%%%%%%%%%%%%%%%%%%%%%%%%

%%%%%%%%%%%%%%%%%%%%%%%%%%%%%%%%%%%%%%%%%%%%%%%%%
\begin{figure}[t]
\includegraphics[width=1.02\linewidth]{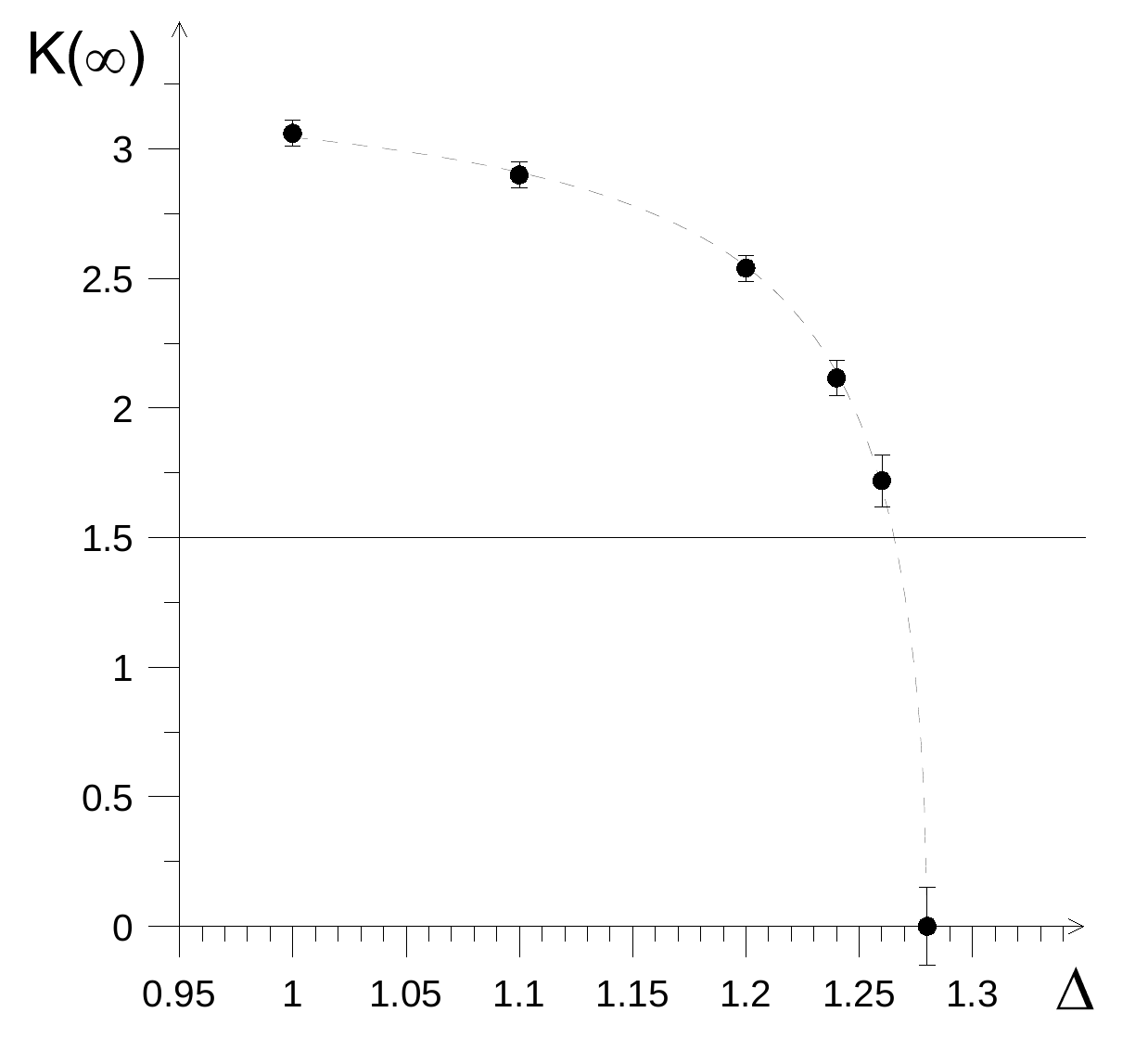}
\caption{\label{fig:KD} The asymptotic  values of the Luttinger parameter extracted from extrapolations of classical flows to $L\to \infty$.
The dashed line is the fit $K(\Delta, L=\infty)=a\sqrt{x/(x+b)}$, $x=\Delta_c^{\rm (cl)}\!-\!\Delta$ ($a$ and $b$ are fitting parameters) consistent with the behavior of $K(\Delta, L=\infty)$ predicted by Eq.~(\ref{eq:flow_ns}). The quantum-critical value of disorder, $\Delta_c=1.265(10)$,
is obtained from the condition $K(\Delta_c, L=\infty)=K_c=3/2$. }
\end{figure}
%%%%%%%%%%%%%%%%%%%%%%%%%%%%%%%%%%%%%%%%%%%%%%%%%

%%%%%%%%%%%%%%%%%%%%%%%%%%%%%%%%%%%%%%%%%%%%%%%%%
\begin{figure}[t]
\includegraphics[width=1.02\linewidth]{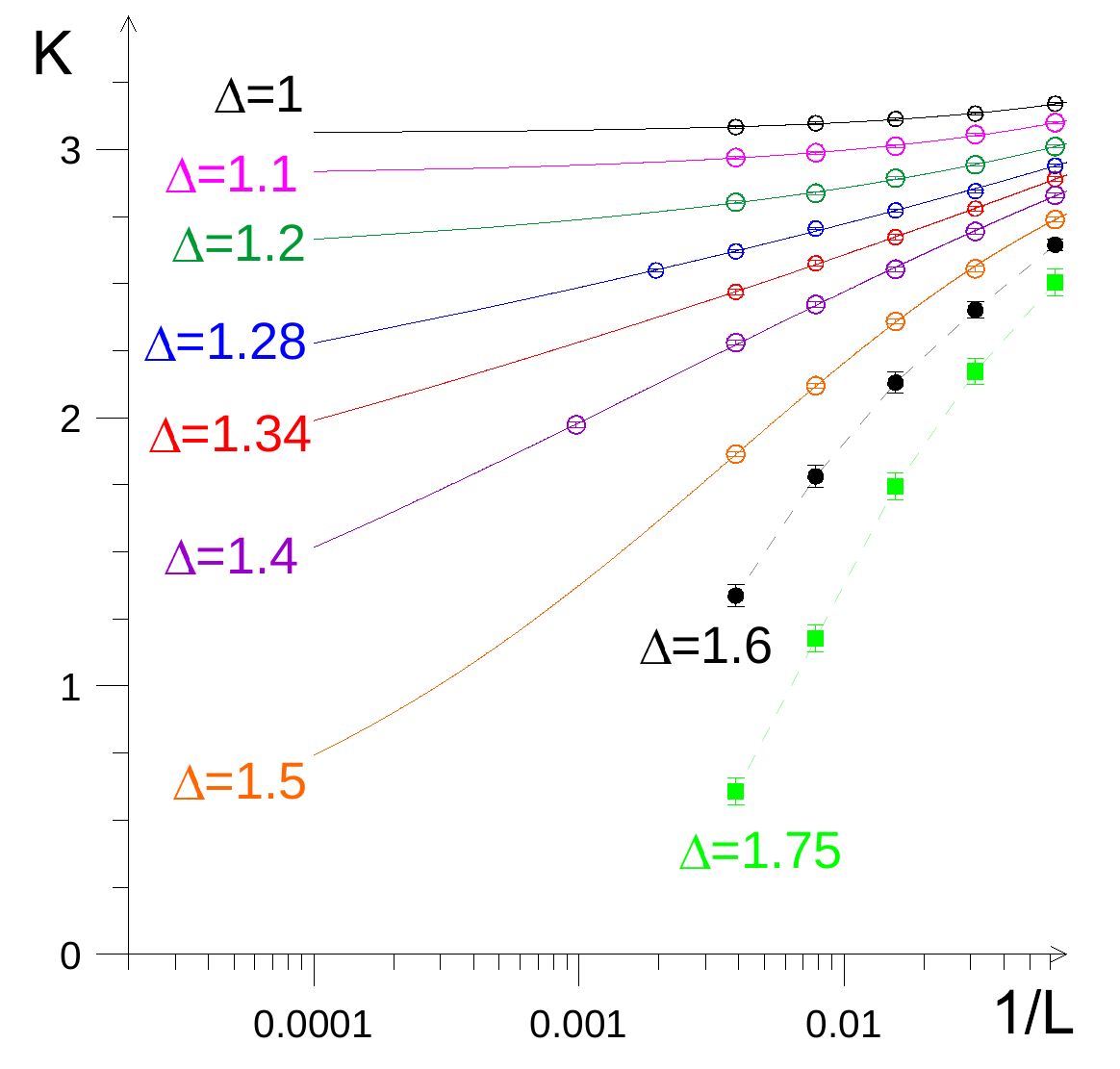}
\caption{\label{fig:DKall} (Color online). The flows of median values of  the Luttinger parameter. In view of a perfect self-averaging (and absence of any appreciable $L$-dependence) of $\kappa$,
the solid lines correspond directly to their counterparts in Fig.~\ref{fig:glory} processed in accordance with Eq.~(\ref{eq:Luttinger}). The dashed lines for the $\Delta=1.6$  and
$\Delta=1.75$ data are added simply to guide an eye (see the discussion in the text).}
\end{figure}
%%%%%%%%%%%%%%%%%%%%%%%%%%%%%%%%%%%%%%%%%%%%%%%%%

%%%%%%%%%%%%%%%%%%%%%%%%
\subsection{Model and Technique}
%%%%%%%%%%%%%%%%%%%%%%%%

We consider the disordered J-current model~\cite{Wallin94},
\begin{equation}
H = \sum_{\bm{n}} t J_{{\bm n}, x}^2 - \vert U \vert J_{{\bm n}, \tau}^2 - (\mu - \epsilon_x)  J_{{\bm n}, \tau}.
\end{equation}
Here, the integer vector $\bm{n} = (x, \tau )$ labels sites of the two-dimensional square space-time lattice of dimension $L_x \times L_{\tau}$,
$J_{{\bm n}, \alpha}$  are integer ``currents"
living on lattice bonds labeled by ${\bm n}, \alpha$ where
$\alpha = x, \tau $ stands for unit vectors pointing in the positive discrete space and imaginary time directions, respectively. The allowed values for the currents are $|J_{{\bm n}, x}|\le 1$ and $ 0 \le J_{{\bm n}, \tau}\le 2$.
(We will write $L_x$ and $L$ intermittently for the system size).
The allowed configurations of bond currents are subject to the zero-divergence constraint
$ \sum_{\alpha} (J_{\bm{n},-\alpha} +J_{\bm{n}, \alpha}) = 0$, where, by definition,
$J_{\bm{n},-\alpha}  = - J_{\bm{n}-\alpha , \alpha}$.
We consider attractive interactions and choose them as the unit, $U = 1$. Disorder $\epsilon_x$ uniformly chosen in the box $[-\Delta, \Delta]$  is added to the non-zero chemical potential, which breaks particle-hole symmetry (the SF-BG transition
is then expected at $K_c = 3/2$ according to the GS scenario). In the absence of disorder, a first order SF-MI transition occurs in a broad parameter range
starting from $(\mu/U = -2, t=0)$. To test the theory on presently achievable system sizes,
we take $t=1/4$ and $\mu = -1.8$ in the vicinity of this first-order transition, and expect strong-disorder fluctuations
to generate local insulating regions responsible for RWLs.

The central (dimensionless) quantity of interest is
the superfluid stiffness, $n_s $.
Also important is the compressibility, $\kappa$,  since a combination
of $\kappa$ and $n_s$  gives the Luttinger parameter $K$,
\begin{equation}
 K = \pi \sqrt{n_s \kappa} \, .
\label{eq:Luttinger}
\end{equation}
The superfluid stiffness and compressibility can be computed from the winding numbers $W_x$ and $W_{\tau}$ in space and time,
respectively:
\begin{equation}
n_s    = \left( \langle W_x^2      \rangle  - \langle W_x    \rangle^2 \right) (L_x / L_{\tau}) \, ,
\label{eq:n_s_W}
\end{equation}
\begin{equation}
\kappa = \left(\langle W_{\tau}^2 \rangle  - \langle W_\tau \rangle^2 \right) ( L_{\tau} / L_x)\, .
\label{eq:kappa_W}
\end{equation}
Here, the angular brackets stand for statistical averaging for a given disorder realization.
Assuming a transition with dynamical exponent $z=1$,
we perform finite size scaling with constant aspect ratio $L_{\tau} / L_x = 1/2$.
In the figures we plot results for winding number fluctuations which explains an extra factor of $1/2$.
Simulations were done by employing the classical worm algorithm~\cite{easyworm}.
We typically consider between 5,000 and 20,000 disorder realizations in order to have converged answers; from a couple of 100 realizations one cannot determine the distribution functions.

%%%%%%%%%%%%%%%%%%%%%%%%
\subsection{The Results}
%%%%%%%%%%%%%%%%%%%%%%%%

Our first observation is that, in accordance with  the theory of Sec.~\ref{sec:2} (as well as with the standard GS scenario),
it is only  the superfluid stiffness that gets substantially renormalized in the critical region and in the insulating phase, while the compressibility remains constant (within error bars) for all system sizes, see Fig.~\ref{fig:sf_kappa}.

In the presence of strong disorder, the distribution of $n_s^{-1}$ becomes quite broad (see App.~\ref{sec:appA} for the full distributions). %\cite{epaps}.
The flows of relative widths as functions of the system size, $R_s(L)$,  are shown in Fig.~\ref{fig:width}.
In the theoretical limit of large $L$, the function $R_s(L)$ is supposed to decrease (tending to zero at $L\to \infty$) in the superfluid phase,
and increase in the insulating phase. As we see in Fig.~\ref{fig:width}, these two distinctively different asymptotic trends set in only at $L\gtrsim100$
(with the separatrix at $\Delta <1.28$). At $L < 100$, the relative width $R_s(L)$ increases on both sides of the separatrix. This, however, does not mean
that  the theory of Sec.~\ref{sec:2} does not apply at $L \lesssim 100$. Actually, the opposite is true since the initial growth of $R_s(L)$ is naturally attributed  to the fact that the function $\tilde{F}(\ln L)$
in Eq.~(\ref{eq:flow_ns}) is small already at $L\sim 1$, rather than only in the asymptotic limit of $L\gg 1$. Indeed, as is clear then from Eq.~(\ref{eq:flow_ns}) [that now applies almost from
$L\sim 1$], for $n_s^{-1}$ to get significantly renormalized---which is a necessary condition for developing substantial sample-to-sample fluctuations---the system size $L$ has to increase dramatically.

Figure~\ref{fig:glory} reveals a remarkable quantitative agreement between the numerical results for the median values of $n_s^{-1}(L)$ and the theory of Sec.~\ref{sec:2}. Our fitting  procedure
is based on Eq.~(\ref{eq:flow_ns}) in which the  function $\tilde{F}(\ln L)$ is chosen to be a ($\Delta$-dependent) constant, $\tilde{F}\equiv B(\Delta)$, treated as a fitting parameter for a given strength of disorder. The second fitting parameter is the exponent $\zeta(\Delta)$. A crucial consistency condition then is that the  functions $B(\Delta)$ and $\zeta(\Delta)$ be structureless in the vicinity of the classical critical value of disorder, meaning that our data reveals nothing beyond Eqs.~(\ref{eq:prob})-(\ref{eq:flow_ns}).  As we see in Figs.~\ref{fig:zeta} and \ref{fig:B}, this condition is met.

The genuine (quantum) critical value of disorder, $\Delta_c=1.265(10)$, is found by extrapolating $K(\Delta, L)$ to $L=\infty$ and requiring $K(\Delta_c, \infty)=K_c=3/2$, see Fig.~\ref{fig:KD}.
This procedure completely ignores the renormalization of $K(L)$ due to instanton pairs, which is justified by the vanishingly small fugacity of the latter, as argued in Sec.~\ref{sec:2}.
The value of $\Delta_c$ has overlapping error bars with the classical critical value $\Delta_c^{\rm (cl)} =1.275(10)$ obtained in  Fig.~\ref{fig:zeta}. This is due to the fact that $K$ is a steep  function of $n_s$ at $n_s\to 0$, which is the reason we can find extrapolated $K(\infty)$ values very close to $K_c=3/2$.  [Our error bars do not account for possible systematic bias originating from the simple choice of  $\tilde{F}\equiv B(\Delta)$. The corresponding analysis, though possible, goes beyond the goals of the present paper.]

Figure~\ref{fig:DKall}, to be compared to Fig.~\ref{fig:GS} and contrasted  to Fig.~\ref{fig:KT}, shows the flows of $K(L)$. The qualitative agreement with Fig.~\ref{fig:GS} and the pronounced difference from Fig.~\ref{fig:KT} are clearly seen. Especially instructive are the data sets for $\Delta=1.6$  and  $\Delta=1.75$. At those values of disorder the system is unquestionably well inside the insulating phase---to the extent that even fitting the flow by the ansatz of Eq.~(\ref{eq:flow_ns}) can hardly be justified because Eq.~(\ref{eq:flow_ns}) is derived under the assumption
of $\zeta \ll 1$, while  $\zeta(\Delta=1.6) \approx 0.8$ (see Fig.~\ref{fig:zeta} ). Nevertheless, the marginal superfluid behavior persists at $\Delta=1.6$  and  $\Delta=1.75$ till system sizes of few hundreds. Most importantly, the flows for $\Delta=1.6$  and  $\Delta=1.75$ go well below the quantum-critical value $K_c=3/2$ without any change of their character. Qualitatively,
the flows for $\Delta=1.6$  and  $\Delta=1.75$ are very similar to the flow for $\Delta=1.5$ extrapolated to $K(L)\sim 1$, which is  dramatically different from
the situation sketched in Fig.~\ref{fig:KT}.

Note that  for $\Delta=1.6$  and  $\Delta=1.75$ marginal superfluidity persists till the
Kane-Fisher point $K_{\rm KF}=1$, in agreement with our conclusion concerning the suppressed fugacity of instanton--anti-instanton pairs. Because for our model the classical separatrix and quantum critical line are so close to each other, there is no chance of seeing a quasi-plateau between $K=1$ and $K=1.5$.
The slow flow thus shows that, on physical length scales, we are dealing with a mesoscopic situation where the notion of the thermodynamic state
is less meaningful than the flow itself: Making cuts at fixed $L$ or fixed $K=2.5$ in Fig.~\ref{fig:DKall} reveals little information in itself but is meaningful when understood in terms of the flow equations. Hence accurately tracing the evolution of $K$ with $L$ is absolutely
crucial for understanding the underlying physics.
A recent numerical study~\cite{Vojta} of a particle-hole symmetric (1+1)-dimensional XY model came to opposite conclusions as the ones presented here. The authors of Ref.~\onlinecite{Vojta} analyzed their data by monitoring the susceptibility, Luttinger parameter, and correlation length $\xi$ and found a non-universal $K_c$ for strong enough disorder. Their type of analysis is however incompatible with our data, but we believe that resolving the flow of $K$ with $L$ on the critical separatrix for strong disorder, which is lacking in the paper, could settle the discrepancy in favor of the theory outlined in this work.

%%%%%%%%%%%%%%%%%%%%%%%%%%%%%%%%%%%
\section{Conclusion}
\label{sec:5}
%%%%%%%%%%%%%%%%%%%%%%%%%%%%%%%%%%
Our work proves that strong disorder may lead to a
logarithmic flow of the stiffness
coming from the classical-field renormalization while compressibility remains constant. Although
the universality class of the superfluid to Bose glass quantum phase transition does not change,
the classical-field mesoscopic physics renders this universality essentially
non-observable on physically relevant length scales, since a peculiar mesoscopic behavior
persists up to extremely large system sizes. Even deep in the insulating side
a physical ({\it e.g.}, cold-atomic) system can demonstrate a marginal superfluid response, with strong sample-to-sample fluctuations.
It makes hence little sense to identify the thermodynamic phase for certain Hamiltonian parameters if only finite samples are available.
We believe that this mesoscopic behavior, illustrated by numerical
data, is the long-sought reconciliation of classical-field criticality with the
established picture of the superfluid-to-glass quantum phase transitions
(which remains intact in the academic thermodynamic limit).
Fitting numerical/experimental data for the superfluid stiffness as a function of disorder and the system size with the  classical-field
renormalization flow allows one to accurately extract the critical value of disorder and predict the behavior of the system at much larger sizes.
In future work, we plan to employ this theory for completing the SF-BG phase diagrams
of the disordered Bose-Hubbard model in the weakly interacting Bose gas regime in
three~\cite{3d_2}, two~\cite{Soyler}, and one dimension.

Finally, we proved that---regardless of the possible divergence of variance of inverse superfluid density---any quantum critical flow possesses self-averaging properties and is thus subject to an asymptotic hydrodynamic description.
In our view, this theorem rules out quantum strong-coupling criticality scenarios.

%%%%%%%%%%%%%%%%%%%%%%%%%%%%%%%%%%%%%%%%%%%%%%%%
We are grateful to Ehud Altman, Thierry Giamarchi, Victor Gurarie, David Pekker,  Susanne Pielawa, and Anatoli Polkovnikov  for valuable discussions.
This work was supported by the National Science Foundation
under the grant PHY-1005543.  LP thanks the hospitality of the Kavli Institute for Theoretical Physics at UCSB and at Beijing, where part of this work was done.\\

%%%%%%%%%%%%%%%%%%%%%%%%%%%%%%%%%%%%%%%%%%%%%%%%%
\begin{figure}[t]
\includegraphics[width=1.02\linewidth]{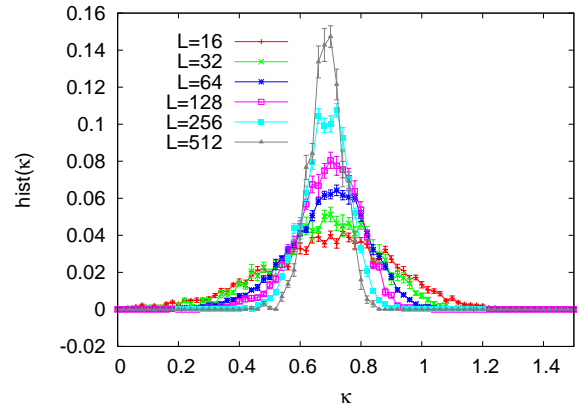}
\caption{\label{fig:distr_kappa_D128} (Color online). Distribution of the compressibility $\kappa$ for $\Delta = 1.28$ close to the critical line.}
\end{figure}
%%%%%%%%%%%%%%%%%%%%%%%%%%%%%%%%%%%%%%%%%%%%%%%%%

%%%%%%%%%%%%%%%%%%%%%%%%%%%%%%%%%%%%%%%%%%%%%%%%%
\begin{figure}[t]
\includegraphics[width=1.02\linewidth]{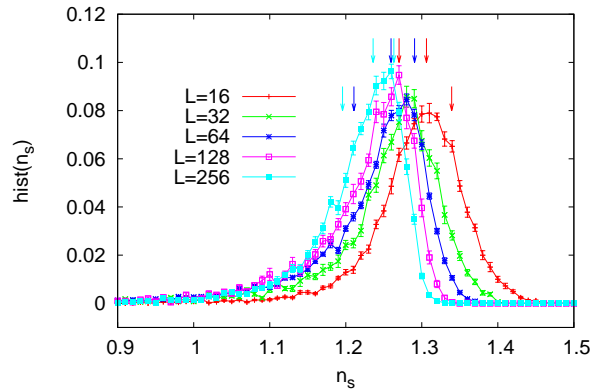}
\caption{\label{fig:distr_sf_D1} (Color online).  Distribution of the stiffness $n_s$ for $\Delta = 1$ in the superfluid phase. Arrows indicate the 25, 50 and 75 percentile, from left to right for $L=16, L=64$ and $L=256$.}
\end{figure}
%%%%%%%%%%%%%%%%%%%%%%%%%%%%%%%%%%%%%%%%%%%%%%%%%

%%%%%%%%%%%%%%%%%%%%%%%%%%%%%%%%%%%%%%%%%%%%%%%%%
\begin{figure}[t]
\includegraphics[width=1.02\linewidth]{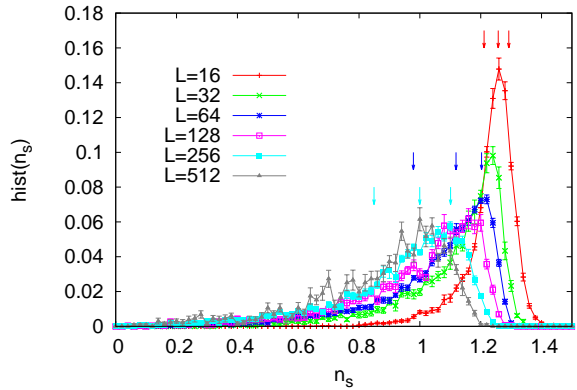}
\caption{\label{fig:distr_sf_D128} (Color online).  Distribution of the stiffness $n_s$ for $\Delta = 1.28$ close to the critical line. Arrows indicate the 25, 50 and 75 percentile, from left to right for $L=16, L=64$ and $L=256$.}
\end{figure}
%%%%%%%%%%%%%%%%%%%%%%%%%%%%%%%%%%%%%%%%%%%%%%%%%

%%%%%%%%%%%%%%%%%%%%%%%%%%%%%%%%%%%%%%%%%%%%%%%%%
\begin{figure}[t]
\includegraphics[width=1.02\linewidth]{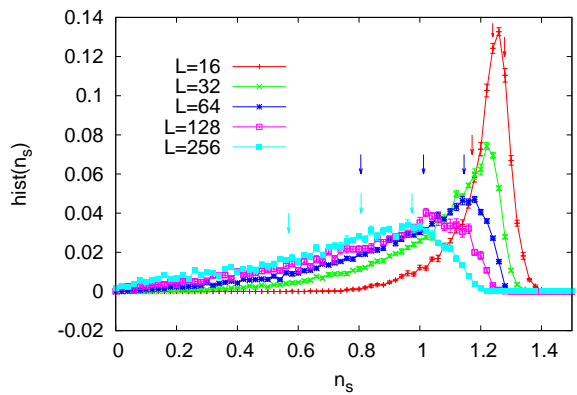}
\caption{\label{fig:distr_sf_D14} (Color online).  Distribution of the stiffness $n_s$ for $\Delta = 1.4$ close to the critical line, but in the insulating regime. Arrows indicate the 25, 50 and 75 percentile, from left to right for $L=16, L=64$ and $L=256$.}
\end{figure}
%%%%%%%%%%%%%%%%%%%%%%%%%%%%%%%%%%%%%%%%%%%%%%%%%

%%%%%%%%%%%%%%%%%%%%%%%%%%%%%%%%%%%%%%%%%%%%%%%%%
\begin{figure}[t]
\includegraphics[width=1.02\linewidth]{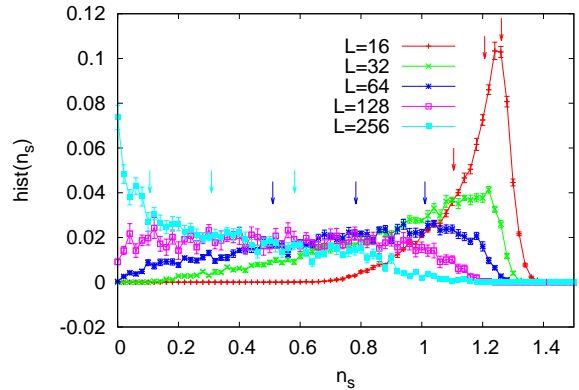}
\caption{\label{fig:distr_sf_D16} (Color online).  Distribution of the stiffness $n_s$ for $\Delta = 1.6$ in the insulating regime. Arrows indicate the 25, 50 and 75 percentile, from left to right for $L=16, L=64$ and $L=256$.}
\end{figure}
%%%%%%%%%%%%%%%%%%%%%%%%%%%%%%%%%%%%%%%%%%%%%%%%%

%%%%%%%%%%%%%%%%%%%%%%%%%%%%%%%%%%%%%%%%%%%%%%%%
\appendix
\section{Full Distribution Functions}
\label{sec:appA}
%%%%%%%%%%%%%%%%%%%%%%%%%%%%%%%%%%

Although it is argued in this work that the median and the (relative) width defined in terms of percentiles of the distribution give a perfect description of the physics, we nevertheless want to show the full distribution functions in this Appendix. On the one hand this gives the reader a better appreciation of the broadness of these distributions (and the corresponding numerical challenge to resolve the distribution) as well as an indication of where the median lies. On the other hand, the strong disorder scenario of Ref.~\onlinecite{Altman2004} puts special emphasis on the tails of the distributions for the stiffness $n_s$ and we also would like to provide this information for completeness.

We start with the distribution for the compressibility. As argued in the text, the compressibility plays no important role in the flow. In Fig.~\ref{fig:distr_kappa_D128} we show the distribution of $\kappa$ for different system sizes at $\Delta=1.28$ which is very close to the critical line. The distributions show strong self-averaging when increasing the system size, as expected. So for $\kappa$ we are able to define the average and the variance without problem.
This picture is not altered in the superfluid or in the insulating phase (not shown). We are thus safe to ignore the renormalization of the compressibility (even though the distribution is very broad for small system sizes).

We proceed with the distributions of the stiffness $n_s$. Deep in the superfluid phase we see with the naked eye in Fig.~\ref{fig:distr_sf_D1} for $\Delta = 1$ that self-averaging occurs even though it is a slow process with system size. The median tracks the maximum of the distribution well (they will coincide in the thermodynamic limit). When we increase $\Delta$ to be close to the critical point (see Fig.~\ref{fig:distr_sf_D128} for $\Delta=1.28$) the median shifts to lower values for small system sizes. However, the flow seems to converge to a finite value in the superfluid phase, while the relative width is decreasing in the superfluid phase and increasing in the insulating phase (see Fig.~\ref{fig:width}). On the numerically accessible system sizes, nothing is seen to occur around $n_s=0$ near the critical line.
For slightly larger $\Delta$ (see Fig.~\ref{fig:distr_sf_D14} for $\Delta=1.4$) the flow to the insulator can nicely be tracked. For small system sizes such as $L=16$ the distribution has a well-developed peak but we note an asymmetric and pronounced tail for smaller $n_s$ that develops strongly when we increase $L$. For $L=256$ the distribution approaches $n_s=0$  in such a way that the variance for $n_s^{-1}$ certainly diverges. After that, the insulating behavior is unstoppable and quickly becomes  apparent in a similar fashion as seen deep in the insulating regime (see Fig.~\ref{fig:distr_sf_D16} for $\Delta=1.6$) already at smaller system sizes ($L=128$ and $L=256$): The distribution quickly broadens and develops a strong overlap with $n_s=0$ for increasing $L$. Ultimately the median will flow to 0.


\begin{thebibliography}{99}
\bibitem{GS} T. Giamarchi and H.J. Schulz, Europhys. Lett. {\bf 3}, 1287 (1987); Phys. Rev. B {\bf 37}, 325 (1988).
\bibitem{Fisher} M.P.A. Fisher, P.B. Weichman, G. Grinstein, and D.S. Fisher, Phys. Rev. B {\bf 40}, 546 (1989).
\bibitem{Weichman} P. B. Weichman, Mod. Phys. Lett. B \textbf{22}, 2623 (2008).
\bibitem{Giamarchi_2001}  T. Giamarchi, P. L. Doussal, and E. Orignac, Phys. Rev. B  {\bf 64}, 245119 (2001).
\bibitem{3d_1}  L. Pollet, N.V. Prokof'ev, B. V. Svistunov, and M. Troyer, Phys. Rev. Lett. {\bf 103}, 140402 (2009).
\bibitem{3d_2} V. Gurarie, L. Pollet, N. V. Prokof'ev, B. V. Svistunov, and M. Troyer, Phys. Rev. B {\bf 80}, 214519 (2009).
\bibitem{Soyler} S.G. Soyler, M. Kiselev, N. V. Prokof'ev, and B. V. Svistunov, Phys. Rev. Lett. {\bf 107}, 185301 (2011).
\bibitem{large_scale_1} F. Alet and E.S. S{\o}rensen, Phys. Rev. E {\bf 67}, 015701 (2003).
\bibitem{large_scale_2} N. Prokof'ev and B. Svistunov,  Phys. Rev. Lett. {\bf 92}, 015703 (2004).

%%%%%%%%%%%% experiment with ultra cold atoms %%%%%%%%%%
\bibitem{exp_weak_1} J. Billy, V. Josse, Z. Zuo, A. Bernard, B. Hambrecht, P. Lugan, D. Cl\'ement1, L. Sanchez-Palencia, P. Bouyer, and A. Aspect, Nature {\bf 453}, 891 (2008).
\bibitem{exp_weak_2} L. Fallani, J. E. Lye, V. Guarrera, C. Fort, and M. Inguscio, Phys. Rev. Lett.  \textbf{98}, 130404 (2007).
\bibitem{exp_weak_3}  G. Roati, C. D'Errico, L. Fallani, M. Fattori, C. Fort, M. Zaccanti, G. Modugno, M. Modugno, and M. Inguscio, Nature {\bf 453}, 895 (2008).
\bibitem{exp_weak_4} B. Deissler, M. Zaccanti, G. Roati, C. D'Errico, M. Fattori, M. Modugno, G. Modugno, and M. Inguscio, Nature {\bf 6}, 354 (2010).
\bibitem{exp_demarco} M. White, M. Pasienski, D. McKay, S.Q. Zhou, D. Ceperley,
and B.  DeMarco,  Phys. Rev. Lett. {\bf 102}, 055301 (2009).
%%%%%%%%%%%%%%%%%%%%%%%%%%%%%%%%%%%%

\bibitem{instanton} V.A. Kashurnikov, A.I. Podlivaev, N.V. Prokof'ev, and B.V. Svistunov, Phys. Rev. B {\bf 53}, 13091 (1996).
\bibitem{Ristivojevic} Z. Ristivojevic, A. Petkovi\'c, P. Le Doussal, and T. Giamarchi, Phys. Rev. Lett. {\bf 109}, 026402  (2012).
\bibitem{alexander} S. Alexander, J. Bernasconi, W.R. Schneider, and R. Orbach, Rev. Mod. Phys. {\bf 53}, 175 (1981).
\bibitem{Altman2004} E. Altman, Y. Kafri, A. Polkovnikov, and G. Refael, Phys. Rev. Lett. {\bf 93}, 150402 (2004); {\it ibid}
%\bibitem{Altman2010} E. Altman, Y. Kafri, A. Polkovnikov, and G. Refael,
Phys. Rev. B {\bf 81}, 174528 (2010).
\bibitem{Balabanyan} K. G. Balabanyan, N. V. Prokof'ev, and B. V. Svistunov, Phys. Rev. Lett. {\bf 95}, 055701 (2005).
\bibitem{Kane-Fisher} C.L. Kane and M.P.A. Fisher, Phys. Rev. Lett. {\bf 68}, 1220 (1992).
\bibitem{Polkovnikov} We are grateful to Anatoly Polkovnikov for pointing this out.
\bibitem{Feller1} W. Feller {\it An Introduction to Probability Theory and Its Applications}, Vol.1 (3rd Edition), Wiley, ISBN 0-471-25708-7  (1968).
\bibitem{Feller2} W. Feller {\it An Introduction to Probability Theory and Its Applications}, Vol.2, Wiley, ISBN 0-471-25709-5  (1971).
\bibitem{Wallin94} M. Wallin, E.S. S{\o}rensen, S.M. Girvin, and A.P. Young, Phys. Rev. B {\bf 49}, 12115 (1994).
\bibitem{easyworm} N. V. Prokof'ev and B. V. Svistunov, Phys. Rev. Lett. {\bf 87}, 160601 (2001).
\bibitem{Vojta} F. Hrahsheh and T. Vojta, Phys. Rev. Lett. {\bf 109}, 265303 (2012).





\end{thebibliography}
\end{document}